\documentclass[aps,showpacs,showkeys]{revtex4}
\usepackage{mathrsfs}
\usepackage{amssymb}
\usepackage{amsmath}
\usepackage{bm}
\usepackage{graphics}
\usepackage{natbib}
\newcommand{\llangle}{\langle\!\langle}
\newcommand{\rrangle}{\rangle\!\rangle}

\newcommand{\sss}[1]{{\scriptscriptstyle #1}}
\begin{document}
\title{Decoupling and antiresonance in a quantum dot chain
with two neighboring dots coupled to both leads}
\author{Yu Han$^{a,b}$}
\author{Weijiang Gong$^{a}$}\email[Corresponding author.
Fax: +086-024-8367-6883; phone: +086-024-8367-8327; Email address: ]
{weijianggong@gmail.com}
\author{Guozhu Wei$^{a,c}$}

\affiliation{
a. College of Sciences, Northeastern University, Shenyang 110004, China \\
b. Department of Physics, Liaoning University, Shenyang 110036, China \\
c. International Center for Material Physics, Acadmia Sinica,
Shenyang 110015, China}
\date{\today}

\begin{abstract}
Electron transport through a quantum dot chain with two neighboring
dots coupled to both leads is theoretically studied. In such a
system, it is found that only for the even-numbered quantum dot
structure with the same-number quantum dots coupled to each
connecting dot, some eigenstates of the quantum dots decouple from
the leads. Namely, all odd eigenstates decouple from the leads in
the absence of magnetic flux, but all even eigenstates will decouple
from the leads when a magnetic flux is introduced. In addition, by
adjusting the magnetic fluxes through any subring, some eigenstates
decouple from one lead but still couple to the other, and then some
new antiresonances occur.
\end{abstract}
 \maketitle

\bigskip

\section{Introduction}
In the past years, electronic transport through quantum-dot(QD)
systems has been extensively studied both experimentally and
theoretically. The atom-like characteristics of a QD, such as the
discrete electron levels and strong electron correlation, manifest
themselves by the experimental observations of Coulomb
blockade\cite{Thomas,Meirav,Sakaki,Zhang}, conductance
oscillation\cite{Kool}, and Kondo
effect\cite{Mahalu,Gores,Cronenwett,Heary} in the electronic
transport process through a QD. Therefore, a single QD is usually
called an artificial atom, and a mutually coupled multi-QD system
can be regarded as an artificial molecule. With the progress of
nanotechnology, it now becomes possible to fabricate a variety of
coupled QD structures with sizes to be smaller than the electron
coherence\cite{Xie,Shailos}. Thereby, more and more recent
experimental investigations focus on the electronic transport
property through coupled QD
systems\cite{Tarucha,Yu,Vidan,Waugh,Amlani}. In comparison with a
single QD, coupled QD systems possess higher freedom in implementing
some functions of quantum devices, such as the QD cellular
automata\cite{Loss1} and solid-state quantum
computation\cite{Loss2,Klein}.
\par
Many experimental and theoretical works have become increasingly
concerned about the electronic transport through various multi-QD
systems\cite{Iye,Sato,Ihn,Wang,Santos,Zheng1,Aligia,Tor}. According
to the previous research, we know that characteristic of the linear
conductance spectrum consists in the conductance peaks reflecting
the eigenlevels of the coupled QDs. Moreover, the zero point of the
conductance, called antiresonance, has also received much attention,
which is interpreted as the destructive quantum interference among
electron waves going through different paths. Typically, in the
structures with one or several quantum dots side-coupled to a main
conducting channel, the antiresonant points coincide with the
eigenenergies of the dangling QDs. Such a theoretical prediction
about antiresonance has been observed experimentally\cite{Iye,Sato},
which stimulated further theoretical interest in this
topic\cite{Wang,Santos,Zheng1,Aligia,Tor}. Based on the properties
of antiresonance, the applications of some QD structure are
proposed\cite{Tor,Liu,Gong1}.
\par
However, for several QD systems\cite{Orellana,Bao}, not all the
eigenlevels appear in the conductance spectrum and the corresponding
eigenlevels are completely localized, which originate from these
quantum states decoupled from the leads. Thereby, the occurrence of
decoupling modifies the quantum interference of the corresponding
structure in the nontrivial way, so that the electron transport
properties are changed. For instance, it can also give rise to the
appearance of antiresonance. Accordingly, such topics have become a
direction of focusing on the electron transport through the coupled
QDs. Motivated by these previous works, we pay attention to the
electron transport through a quantum dot chain, in which both the
leads couple to two neighboring dots of the chain. As a result, only
for the even-dot structure with the same-number quantum dots
coupling to the connecting dots, some of the eigenstates of the
coupled QDs decouple from the leads. Namely, in the absence of
magnetic flux only the odd eigenstates decouple from the leads,
while the even eigenstates will decouple from the leads when
magnetic flux is introduced. But the antiresonance in the
conductance spectrum is irrelevant to the tuning of magnetic flux.
In addition, by adjusting the magnetic fluxes through each subring,
one can achieve the decoupling of some eigenstates from one lead but
still coupled to the other, which causes the occurrence of new
antiresonance.

\section{model\label{theory}}
The coupled-QD structure we consider is illustrated in
Fig.\ref{structure}(a). The Hamiltonian to describe the electronic
motion in such a structure reads
\begin{equation}
H=H_{C}+H_{D}+H_{T}.   \label{1}
\end{equation}
The first term is the Hamiltonian for the noninteracting electrons
in the two leads:
\begin{equation}
H_{C}=\underset{k\alpha\in L,R}{\sum }\varepsilon
_{k\alpha}c_{k\alpha\sigma}^\dag c_{k\alpha\sigma},\label{2}
\end{equation}
where $c_{k\alpha\sigma}^\dag$ $( c_{k\alpha\sigma})$ is an operator
to create (annihilate) an electron of the continuous state
$|k,\sigma\rangle$ in lead-$\alpha$ with $\sigma$ being the spin
index, and $\varepsilon _{k\alpha}$ is the corresponding
single-particle energy. The second term describes electron in the QD
chain. It takes a form as
\begin{equation}
H_{D}=\sum_{m=1}^\sss{N}\varepsilon _{m}d_{m\sigma}^\dag
d_{m\sigma}+\sum_{m=1}^{N-1}(t_md^\dag_{m+1\sigma}d_{m\sigma}
+{\mathrm {H.c.}}),\label{3}
\end{equation}
where $d^{\dag}_{m\sigma}$ $(d_{m\sigma})$ is the creation
(annihilation) operator of electron in QD-$m$, $\varepsilon_j$
denotes the electron level in the corresponding QD. We assume that
only one level is relevant in each QD and the value of $\varepsilon
_{m}$ is independent of $\varepsilon _{m}=\varepsilon_0$. When the
sequence numbers of the two neighboring dots coupled to both leads
are taken as $j$ and $j+1$, the last term in the Hamiltonian
describes the electron tunneling between the leads and QDs. It is
given by
\begin{equation}
H_{T} =\underset{k\alpha\sigma}{\sum }( V_{\alpha j}d_{j\sigma}^\dag
c_{k\alpha\sigma}+V_{\alpha j+1}d_{j+1\sigma}^\dag
c_{k\alpha\sigma}+{\mathrm {H.c.}}), \label{4}
\end{equation}
where $V_{\alpha j}$ and $V_{\alpha j+1}$ with $\alpha=L, R$ denotes
the QD-lead coupling strength. We adopt a symmetric QD-lead coupling
configuration which gives that $V_{Lj}=Ve^{i\phi_L/2}$,
$V_{Lj+1}=Ve^{-i\phi_L/2}$, $V_{Rj}=Ve^{-i\phi_R/2}$, and
$V_{Rj+1}=Ve^{i\phi_R/2}$ with $V$ being the dot-lead coupling
strength. The phase shift $\phi_\alpha$ is associated with the
magnetic flux $\Phi_\alpha$ threading the system by a relation
$\phi_\alpha=2\pi\Phi_\alpha/\Phi_{0}$, in which $\Phi_{0}=h/e$ is
the flux quantum.
\par
To study the electronic transport properties, the linear conductance
of the noninteracting system at zero temperature is obtained by the
Landauer-B\"{u}ttiker formula
\begin{equation}
\mathcal {G}=\frac{e^{2}}{h}\sum_\sigma
T_\sigma(\omega)|_{\omega=\varepsilon_F},\label{conductance}
\end{equation}
$T(\omega)$ is the transmission function, in terms of Green function
which takes the form as\cite{Meir1,Jauho}
\begin{equation}
T_\sigma(\omega)=\mathrm
{Tr}[\Gamma^LG^r_\sigma(\omega)\Gamma^RG^a_\sigma(\omega)],\label{transmission}
\end{equation}
where $\Gamma^L$ is a $N\times N$ matrix, describing the coupling
strength between the two connecting QDs and the left lead. It is
defined as $[\Gamma^{L}]_{mm'}=2\pi
V_{\sss{L}m}V^*_{\sss{L}m'}\rho_\sss{L}(\omega)$. We will ignore the
$\omega$-dependence of $\Gamma^{L}_{mm'}$ since the electron density
of states in the left lead, $\rho_\sss{L}(\omega)$, can be usually
viewed as a constant. By the same token, we can define
$[\Gamma^R]_{mm'}$. In fact, one can readily show that
$[\Gamma^L]_{mm}=[\Gamma^R]_{mm}$ in the case of identical QD-lead
coupling. Hence we take $\Gamma=[\Gamma^L]_{mm}=[\Gamma^R]_{mm}$ to
denote the QD-lead coupling function. In Eq. (\ref{transmission})
the retarded and advanced Green functions in Fourier space are
involved. They are defined as follows:
$G_{mm',\sigma}^r(t)=-i\theta(t)\langle\{d_{m\sigma}(t),d_{m'\sigma}^\dag\}\rangle$
and
$G_{mm',\sigma}^a(t)=i\theta(-t)\langle\{d_{m\sigma}(t),d_{m'\sigma}^\dag\}\rangle$,
where $\theta(x)$ is the step function. The Fourier transforms of
the Green functions can be performed via
$G_{mm',\sigma}^{r(a)}(\omega)=\int^{\infty}_{-\infty}
G_{mm',\sigma}^{r(a)}(t)e^{i\omega t}dt$. These Green functions can
be solved by means of the equation-of-motion method. By a
straightforward derivation, we obtain the retarded Green functions
which are written in a matrix form as
\begin{eqnarray}
G^r_\sigma(\omega)=\left[\begin{array}{cccccc}
\ddots&& & & &\vdots\\
-t_{j-2}& g_{j-1\sigma}(z)^{-1}&-t_{j-1}&0 & 0&0\\
0&-t_{j-1}&g_{j\sigma}(z)^{-1} & -t_j+i\Gamma_{j,j+1}&0&0\\
0&0& -t_j+i\Gamma_{j+1,j}& g_{j+1\sigma}(z)^{-1}&-t_{j+1}&0\\
0&0&0& -t_{j+1}& g_{j+2\sigma}(z)^{-1}&-t_{j+2}\\
\vdots&& & & &\ddots\\
\end{array}\right]^{-1}\ \label{green},
\end{eqnarray}
with $z=\omega+i0^+$,
$g_{m\sigma}(z)=(z-\varepsilon_{m}+i\Gamma_{mm})^{-1}$, being the
zero-order Green function of the QD-$m$ unperturbed by another QD,
and $\Gamma_{mm'}={1 \over 2}([\Gamma^L]_{mm'}+[\Gamma^R]_{mm'})$.
In addition, the advanced Green function can be readily obtained via
a relation $G^a_\sigma(\omega)=[G^r_\sigma(\omega)]^\dag$.

\par
Notice that the linear conductance spectrum of the coupled QD
structure reflects the eigenenergy spectrum of the ``molecule" made
up of the coupled QDs. In other words, each resonant peak in the
conductance spectrum represents an eigenenergy of the total QD
molecule, rather than the levels of the individual QDs. Therefore,
it is necessary to transform the Hamiltonian into the molecular
orbital picture of the QDs. It is quite helpful to analyze the
numerical results for the linear conductance spectrum, as follows.
We then introduce the electron creation(annihilation) operators
corresponding to the molecular orbits, i.e., $f_{m\sigma}^\dag\;
(f_{m\sigma})$. By the diagonalization of the single-particle
Hamiltonian of the QDs, we find the relation between the molecular
and atomic pictures (here each QD is regarded as an ``atom"). This
is expressed as
$[\bm{f}_\sigma^\dag]=[\bm{\eta}][\bm{d}_\sigma^\dag]$. The $N\times
N$ transfer matrix $[\bm{\eta}]$ consists of the eigenvectors of the
QD Hamiltonian. In the molecular orbital picture, the
single-particle Hamiltonian takes the form: ${\cal
H}=\underset{k\sigma\alpha\in L,R}{\sum }\varepsilon _{\alpha
k}c_{\alpha k\sigma }^\dag c_{\alpha k\sigma}+\sum_{m=1, \sigma}e
_{m}f_{m\sigma}^\dag f_{m\sigma}+\underset{\alpha k\sigma}{\sum }
v_{\alpha m}f_{m\sigma}^\dag c_{\alpha k\sigma}+{\mathrm {h.c.}}$,
in which $e_m$ is the eigenenergy of the QDs; $v_{\alpha
m}=\eta_{mj}V_{\alpha j}+\eta_{m,j+1}V_{\alpha j+1}$, denotes the
coupling between the eigenstate $e_m$ and $|k,\sigma\rangle$ in
lead-$\alpha$. In the molecular orbital picture the retarded Green
function is defined as ${\cal G}^r_{mm',\sigma}=\llangle
f_{m\sigma}|f^\dag_{m'\sigma}\rrangle$. We can define
$\gamma^\alpha_{mn}=2\pi v_{\alpha m} v^*_{\alpha
m'}\rho_\alpha(\omega)$ which denotes the coupling coefficient
between the eigenstate $e_m$ and the leads, and their diagonal
elements are given by
\begin{eqnarray}
&&\gamma^\alpha_{mm}=|\eta_{mj}\sqrt{\Gamma^\alpha_{jj}}
+\eta_{m,j+1}\sqrt{\Gamma^\alpha_{j+1,j+1}}e^{i\phi_\alpha}|^2.\label{gamma}
\end{eqnarray}

\par
It is known that for the $N$-QD structure with $t_m=t_0$ and
$\varepsilon_m=\varepsilon_0$, the eigenenergies are given by
$e_m=\varepsilon_0-2t_0\cos(\frac{m\pi}{N+1})$ and the $[\bm\eta]$
matrix is expressed as
\begin{eqnarray}
&&[\bm\eta]=\notag\\&& \sqrt{\frac{2}{N+1}}\left[\begin{array}{cccc}
\sin\frac{N^2\pi}{N+1} & \sin\frac{N(N-1)\pi}{N+1}&\cdots & \sin\frac{N\pi}{N+1}\\
\sin\frac{N(N-1)\pi}{N+1} & \sin\frac{(N-1)^2\pi}{N+1}&\cdots &\sin\frac{(N-1)\pi}{N+1}\\
\vdots & &&\vdots\\
\sin\frac{N\pi}{N+1}&\sin\frac{(N-1)\pi}{N+1}& \cdots&\sin\frac{\pi}{N+1}\\
\end{array}\right].\ \notag
\end{eqnarray}
Hence we can see that only in the case of $\phi_\alpha=n\pi$, it is
possible for $\gamma^\alpha$ to be equal to zero so that the
corresponding eigenstate decouples from lead-$\alpha$.

\section{Numerical results and discussions \label{result2}}

With the formulation developed in the previous section, we can
perform the numerical calculation to investigate the linear
conductance spectrum of this varietal parallel double QD structure,
namely, to calculate the conductance as a function of the incident
electron energy. Prior to the calculation, we need introduce a
parameter $t_0$ as the units of energy.
\par
We choose the parameter values $t_{m}=\Gamma=t_{0}$ for the QDs to
carry out the numerical calculation. And $\varepsilon_{0}$, the QD
level, can be shift with respect to the Fermi level by adjusting
gate voltage experimentally. Figure \ref{QD2} shows the linear
conductance spectra ($\cal{G}$ versus $\varepsilon_0$) for several
structures with the QD number $N=2$ to 4. It is obvious that the
$N=2$ structure just corresponds to the parallel double QDs with
interdot coupling, which has been mentioned in previous works. Its
conductance spectrum presents a Breit-Wigner lineshape in the
absence of magnetic flux, as shown in Fig.2(a). Such a result can be
analyzed in the molecular orbital representation. Here the
$[\bm\eta]$ matrix, denoting the relation between the molecular and
`atomic' representations, takes a form as $[\bm\eta]={1
\over\sqrt{2}}\left[\begin{array}{cc}
-1& 1\\
1& 1\\
\end{array}\right]$. Then with the help of Eq.
(\ref{gamma}) one can find that here the bonding state completely
decouples from the leads and only the antibonding state couples to
the leads, which leads to the appearance of the Breit-Wigner
lineshape in the conductance spectrum. Besides, introducing the
magnetic flux with $\phi_L=\phi_R=\phi=\pi$ (Hereafter we employ
$\phi$ to denote the magnetic flux through each subring for the
$\phi_L=\phi_R$ case) can change the decoupling state, as exhibited
by the dashed line in Fig.\ref{QD2}(a). In such a case, only the
bonding state couples to the leads and the conductance profile also
shows a Breit-Wigner lineshape.

\par
In Fig.\ref{QD2}(b) the conductance curves as a function of gate
voltage are shown for the 3-QD structure. Obviously, there exist
three conductance peaks in the conductance profiles and no
decoupling quantum state appears. We can clarify this result by
calculating
$\gamma^\alpha_{mm}=\Gamma|\eta_{m1}+\eta_{m2}e^{i\phi}|^2$, and it
is obvious that $\gamma^\alpha_{mm}$ is impossible to be equal to
zero in such a structure despite the adjustment of magnetic flux.
Thus one can not find the decoupling eigenstates, the state-lead
coupling may be relatively weak, though. Just as shown in
Fig.\ref{QD2}(b), in the absence of magnetic flux the distinct
difference of the couplings between the quantum states and leads
offer the `more' and `less' resonant channels for the quantum
interference. Then the Fano effect occurs and the conductance
profile presents an asymmetry lineshape. In addition, the Fano
lineshape in the conductance spectrum is reversed by tuning the
magnetic flux to $\phi=\pi$, due to the modulation of magnetic flux
on $\gamma^\alpha_{mm}$.

\par
When the QD number increases to $N=4$, there will be two
configurations corresponding to this structure, i.e, the cases of
$j=1$ and $j=2$. As a consequence, the conductance spectra of the
two structures remarkably differ from each other. With respect to
the configuration of $j=1$, the electron transport properties
presented by the conductance spectra are similar to those in the
case of the 3-QD structure, as shown in Fig.\ref{QD2}(c), and there
is also no occurrence of decoupling states. However, as for the case
of $j=2$, it is clear that in the absence of magnetic flux, there
are two conductance peaks in the conductance spectrum, which means
that the decoupling phenomenon comes into being. Alternatively, in
the case of $\phi=\pi$, there also exist two peaks in the
conductance profile. But the conductance peaks in the two cases of
$\phi=0$ and $\pi$ do not coincide with one another. We can
therefore find that in this structure, when $\phi=n\pi$ decoupling
phenomena will come up and the adjustment of magnetic flux can
effectively change the appearance of decoupling states. By a
calculation and focusing on the conductance spectra, we can
anticipate that in the case of $\phi=2n\pi$, the odd (first and
third) eigenstates decouple from the leads; In contrast, the even
(second and fourth) eigenstates of the QDs will decouple from the
leads if $\phi=(2n-1)\pi$. In addition, as shown in
Fig.\ref{QD2}(d), the conductance encounters its zero when the level
of QDs is the same as the Fermi level of the system. Furthermore,
the conductance zero, usually called antiresonance, is irrelevant to
the tuning of magnetic flux, even though the changes of couplings
between the eigenstates and leads.

\par
In order to obtain a clear physics picture about decoupling, we
analyze this problem in the molecular orbital representation. By
solving the $[\bm\eta]$ matrix and using the relation
$\gamma^\alpha_{mm}=\Gamma|\eta_{m2}+\eta_{m3}e^{i\phi}|^2$, it is
easy to find in the case of zero magnetic flux, $\gamma^\alpha_{11}$
and $\gamma^\alpha_{33}$ is always equal to zero, which brings out
the completely decoupling of the odd eigenstates from the leads.
Opposite to this case, when $\phi=\pi$ the values of
$\gamma^\alpha_{22}$ and $\gamma^\alpha_{44}$ are fixed at zero. And
such a result leads to the even eigenstates to decouple from the
leads. However, the underlying physics responsible for antiresonance
is desirable to clarify. We then analyze the electron transmission
in the molecular representation. We take the case of $\phi=0$ as an
example, where only two quantum states $e_2$ and $e_4$ couple to the
leads due to decoupling. Accordingly, $e_2$ and $e_4$ might be
called as well the bonding and antibonding states. As is known, the
molecular orbits of coupled double QD structures, e.g, the
well-known T-shaped QDs, are regarded as the bonding and antibonding
states. Therefore, by employing the representation transformation
$[\bm{a}_\sigma^\dag]=[\bm{\beta}][\bm{f}_\sigma^\dag]$, such a
configuration can be changed into the T-shaped double-QD system of
the Hamiltonian $H=\underset{k\sigma\alpha\in L,R}{\sum }\varepsilon
_{\alpha k}c_{\alpha k\sigma }^\dag c_{\alpha
k\sigma}+\sum^2_{\sigma, n=1}\epsilon _{n}a_{n\sigma}^\dag
a_{n\sigma} +ta_{2\sigma}^\dag a_{1\sigma}+\underset{\alpha
k\sigma}{\sum } w_{\alpha1}a_{1\sigma}^\dag c_{\alpha
k\sigma}+h.c.$. By a further derivation, the relations between the
structure parameters of the two QD configurations can be obtained
with $\epsilon_1=\varepsilon_0+t_0$, $\epsilon_2=\varepsilon_0$,
$t=t_0$, and $w_{\alpha1}=V_{\alpha1}$ respectively. Accordingly, we
have $\gamma^\alpha_{22}=\Gamma|\beta_{11}|^2$, and
$\gamma^\alpha_{44}=\Gamma|\beta_{21}|^2$ with
$[\bm\beta]={1\over\sqrt{2\sqrt{5}}}\left[\begin{array}{ccc}
-\sqrt{\sqrt{5}-1} & 2\over \sqrt{\sqrt{5}-1}\\
\sqrt{\sqrt{5}+1}& 2\over \sqrt{\sqrt{5}+1} \\
\end{array}\right].$
The 4-QD structure is then transformed into the T-shaped double QDs
with $\varepsilon_0$ being the level of dangling QD. Just as
discussed in the previous works\cite{Gong1}, in the T-shaped QDs
antiresonance always occurs when the dangling QD level is aligned
with the Fermi level of the system. With the help of such an
analysis, one can then understand that in this 4-QD system, the
antiresonant point in the conductance spectrum is consistent with
$\varepsilon_0=0$.
\par
The occurrence of antiresonance in the T-shaped QDs can be
interpreted as the quantum interference between two kinds of
transmission paths. We demonstrate this issue by rewriting the
Hamiltonian of the T-shaped double QD structure as $H={\cal
H}_{0}+{\cal H}_{t}$ in which
\begin{eqnarray}
{\cal H}_{0} &=& \underset{k \sigma}{\sum }\xi_{Lk
}\alpha_{Lk\sigma}^\dag \alpha_{Lk\sigma}+\underset{kk'\sigma }{\sum
}(t_{kk'}\alpha_{Lk\sigma}^\dag c_{Rk'\sigma}+h.c.)\notag\\&&
+\underset{k \sigma}{\sum }\varepsilon_{Rk }\alpha_{Rk\sigma}^\dag
c_{Rk\sigma}+\sum_\sigma\varepsilon_{2}a_{2\sigma}^{\dag}a_{2\sigma},\notag\\
{\cal H}_{t} &=&\underset{k\sigma}{\sum }\mathcal
{V}_k\alpha_{Lk\sigma}^\dag a_{2\sigma}+h.c..
\end{eqnarray}
Here the old operators $a_{1\sigma}$ and $c_{Lk\sigma}$ are expanded
in terms of this new set: $a_{1\sigma}=\underset{k }{\sum
}\nu_{k}\alpha_{Lk\sigma}$ and $c_{Lk\sigma}=\underset{k ^{'}}{\sum
}\eta_{k,k^{'}}\alpha_{Lk^{'}\sigma}$. Under this new
representation, the electron transmission paths are well described:
One is an electron transmission path whereby the electron starts
from the left lead tunnels directly into the right lead, the other
is a different transmission path from the above one in that the
electron must visit the dangling QD as it tunnels through the QD
structure. Note that the electron visiting the dangling QD will
result in its phase change, and the phase difference between the two
kinds of paths gives rise to the destructive quantum interference.
In Ref.\cite{Liu2}, a detailed discussion is presented.
\par
When paying attention to the $[\bm\eta]$ matrix, one will see that
$\eta_{12}=\eta_{43}$, $\eta_{22}=-\eta_{33}$,
$\eta_{32}=\eta_{23}$, and $\eta_{42}=-\eta_{13}$ for the 4-QD
structure. As a result, such relations give rise to
$\gamma^\alpha_{22}|_{\phi=0}=\gamma^\alpha_{33}|_{\phi=\pi}$ and
$\gamma^\alpha_{44}|_{\phi=0}=\gamma^\alpha_{11}|_{\phi=\pi}$. So,
when $\phi=\pi$ the magnetic flux reverses the lineshape of the
conductance spectrum in the case of $\phi=0$. Based on these
properties, we can realize that the quantum interference in this
case is similar to that in the case of $\phi=0$. Therefore, the
antiresonant point in the conductance spectra is independent of the
adjustment of magnetic flux.

\par
By virtue of Eq.(\ref{gamma}) we can find that in the situation of
$\phi_\alpha=n\pi$ and $\phi_{\alpha'}\neq n\pi$, some eigenstates
of the QDs will decouple from lead-$\alpha$ but they still couple to
lead-$\alpha'$. Figure \ref{phi1} shows the research on electron
transport within the 2-QD and 4-QD $j=2$ structures. For the 2-QD
structure, it is obvious that in the case of $\phi_L=0$ and
$\phi_R=0.5\pi$ both the bonding and antibonding states couple to
lead-R but the bonding state decouples from lead-L. The numerical
result is shown in Fig.\ref{phi1}(a). Clearly, due to decoupling
antiresonance occurs at the point of $\varepsilon_0=t_0$. This
indicates that in such a case antiresonance will come into being
when the decoupling state is tuned to be consistent with the Fermi
level. As discussed in our previous works, this antiresonance arises
from the destructive interference among electron going through two
kinds of transmission paths. Then in the case of $\phi_L=0$ and
$\phi_R=\pi$, the bonding state decouples from lead-$L$ with the
antibonding state decoupling from lead-R. So in this case there is
no channel for the electron tunneling and the conductance is always
equal to zero, despite the shift of gate voltage. On the other hand,
by fixing $\phi_R=\pi$ and increasing $\phi_L$ to $0.5\pi$, one will
see that the decoupling of antibonding state from lead-R results in
the antiresonance at the point of $\varepsilon_0=-t_0$. With regard
to the 4-QD with $j=2$ structure, the decoupling-induced
antiresonance is also remarkable in the case of $\phi_\alpha=n\pi$
and $\phi_{\alpha'}\neq n\pi$. As shown in Fig.\ref{phi1}(b) there
are two kinds of antiresonant points in such a case: one originates
from the quantum interference between the connecting eigenstates,
and the other is caused by the decoupling states.
\par
Based on the above analysis, we can expect that in the structure of
with $t_m=t_0$ and $\varepsilon_m=\varepsilon_0$, when $N$ is even
and $j=\frac{N}{2}$ there must be the appearance of decoupling
eigenstates. This expectation can be confirmed because of
$\eta_{mj}=-\eta_{m,j+1} (m\in \text{odd})$ and $
\eta_{mj}=\eta_{m,j+1} (m\in \text{even})$ for the QDs. The
numerical results in Fig.\ref{6-8}, describing the conductances of
6-QD and 8-QD structures, can support our conclusion. Besides, with
the help of the representation transformation the positions of
antiresonance can be clarified by transforming these structure into
the T-shaped QD systems.

\par
In Fig.\ref{semi1} the linear conductances of the semi-infinite and
infinite QD chains are presented as a function of gate voltage. As
shown in Fig.\ref{semi1}(a)-(b), for the case of semi-infinite QD
chains with $j=1$ and 2, there is no conductance peaks consistent
with any eigenlevel since the eigenstates of the QDs become a
continuum in such a case. It should be pointed out that although
some eigenstates decouple from the leads, it can not affect the
electron transport since the electron transmission paths can not be
differentiated in the case of continuum. Thus no antiresonance
appears in the conductance spectra. With this point of view, it is
easy to understand that for the infinite QD chain, there is also no
antiresonance in the conductance profile, as shown in
Fig.\ref{semi1}(c). However, when investigating the influence of the
difference between $\phi_L$ and $\phi_R$ on the electron transport,
we find that in the situation of $\phi_L=0$ and $\phi_R=0.5\pi$ the
conductance of the infinite QD chain encounters its zero at the down
side of energy band, as shown in Fig.\ref{semi2}(b). Such a result
can be explained as follows. The coupling of a semi-infinite QDs to
QD-$j$ indeed brings out an additional self-energy to its level,
which can be written out explicitly as
$\Sigma=\frac{1}{2}(-\varepsilon_0-i\sqrt{4t_0^2-\varepsilon_0^2})$,
which renormalizes the level of QD-$j$ $\varepsilon_0$ to
$\varepsilon_0+\Sigma$. At the upper side of the energy band, the
renormalized QD level becomes $\frac{\varepsilon_0}{2}$ because of
$\Sigma=-\frac{\varepsilon_0}{2}$ here. Similarly, the level of
QD-($j$+1) is equal to $\frac{\varepsilon_0}{2}$. Therefore, at this
point the structure is transformed into the double-dot
configuration. Based on our discussion on the quantum interference
of the double-dot system, we know that when the bonding state
$e_1=\frac{\varepsilon_0}{2}-t_0$ is aligned with the Fermi level,
the linear conductance presents antiresonance. With these knowledge,
one can clarify the occurrence of conductance zero in such a case.
Alternatively, a similar reason gives rise to antiresonance at the
up side of the energy band for the case of $\phi_L=0.5\pi$ and
$\phi_R=\pi$. In addition, it is apparent that in the case of
$\phi_L=0$ and $\phi_R=\pi$ the conductance is always fixed at zero.
This result can be readily explained that in such a situation any
eigenstate coupled to lead-$\alpha$ is inevitable to decouple from
lead-$\alpha'$, though the eigenstates of the QDs is continuum.
Thus, there is still no channels for the electron transport.
\par
Before concluding, we have to make a remark regarding the many-body
effect which we have by far ignored. As is known, the many-body
effect is an important origin for the peculiar transport properties
in QDs. Usually, the many-body effect is incorporated by considering
only the intradot Coulomb repulsion, i.e., the Hubbard term. If the
Hubbard interaction is not very strong, we can truncate the
equations of motion of the Green functions to the second order. By a
straightforward derivation, we find that in such an approximation,
the Green function is redefined as
\begin{equation}
g_{m\sigma}(z)=\big[\frac{z-\varepsilon_m}{1+\frac{U_m\langle
n_{m\bar{\sigma}}\rangle}{z-\varepsilon_m-U_m}}+i\Gamma_{mm}\big]^{-1}.
\end{equation}
It can be expected that the conductance spectrum will be split into
two groups\cite{refChen,refHZheng1,refHZheng2}, and in each group
the electron transport properties in the noninteracting case will
remain.

\section{summary}
With the help of nonequilibrium Green function technique, the
electron transport through a QD chain is theoretically studied. In
such a system both the leads couple to the two neighboring QDs of
the chain. It has been found that only for the even-numbered QD
structure with the same-number QDs coupling to the connecting QDs,
some eigenstates of such coupled QDs decouple from the leads. To be
concrete, in the absence of magnetic flux the odd eigenstates of the
QDs decouple from the leads, whereas its even eigenstates decouple
from the leads when an appropriate magnetic flux is introduced. In
addition, the antiresonance in the conductance spectra is irrelevant
to the tuning of magnetic flux. By means of the representation
transformation, such phenomena were analyzed in detail. By adjusting
the magnetic fluxes through each subring, we found that some
eigenstates of these QDs decoupled from one lead but still coupled
to the other, which causes the occurrence of new antiresonance.
These results vanish for the case of the infinite QD chain.

\clearpage

\bigskip
\begin{figure}
\caption{Schematic of the QD chain with two neighboring QDs coupled
to both leads. Two magnetic fluxes $\Phi_L$ and $\Phi_R$ thread the
subrings in the structure.\label{structure}}
\end{figure}

\begin{figure}
 \caption{The linear
conductance spectra of N-QD chains with $N=2$ to 4. The structure
parameters take the values as $\Gamma=t_{m}=t_{0}$, with $t_0$ being
the unit of energy. \label{QD2}}
\end{figure}

\begin{figure}
 \caption{The calculated
conductance spectra of the 2-QD and 4-QD structures by the presence
of $\phi_\alpha=n\pi$ and $\phi_{\alpha'}\neq n\pi$.
 \label{phi1}}
\end{figure}

\begin{figure}
\caption{(a) The conductances of 6-QD system with $j=3$. (b) The
conductances of 8-QD structure in the case of $j=4$. \label{6-8}}
\end{figure}

\begin{figure}
 \caption{The conductances
of the semi-infinite and infinite QD chains. \label{semi1}}

\end{figure}

\begin{figure} \caption{The conductances
of the semi-infinite and infinite QD chains in the presence of
$\phi_\alpha=n\pi$ and $\phi_{\alpha'}\neq n\pi$. \label{semi2}}
\end{figure}

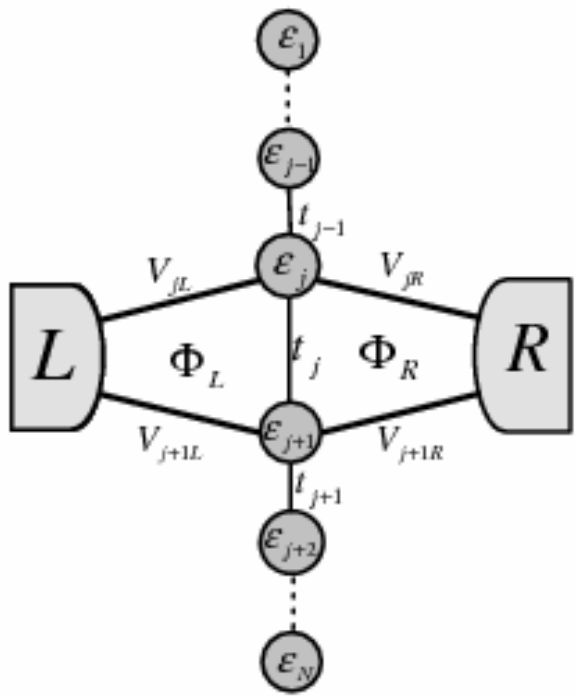
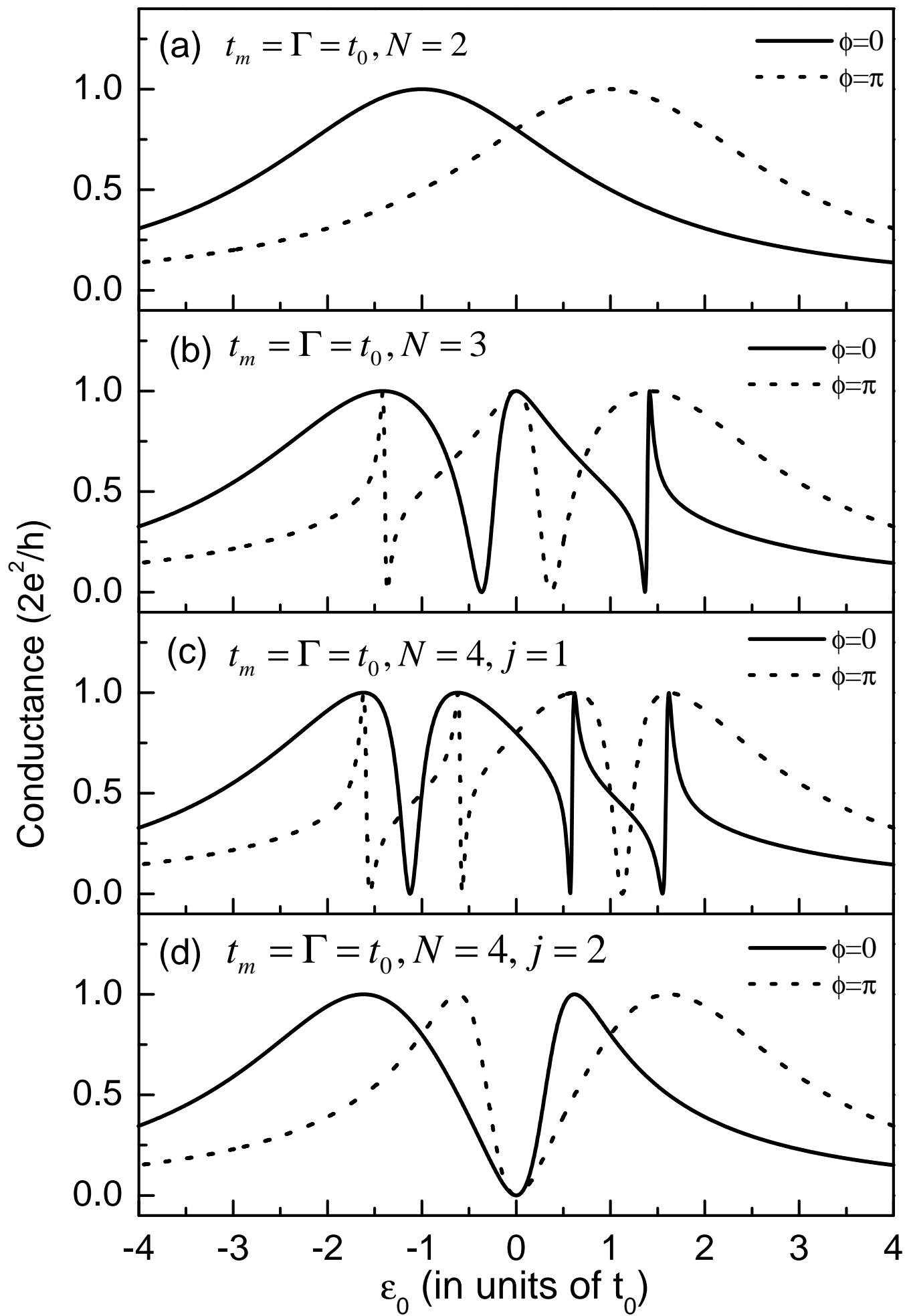
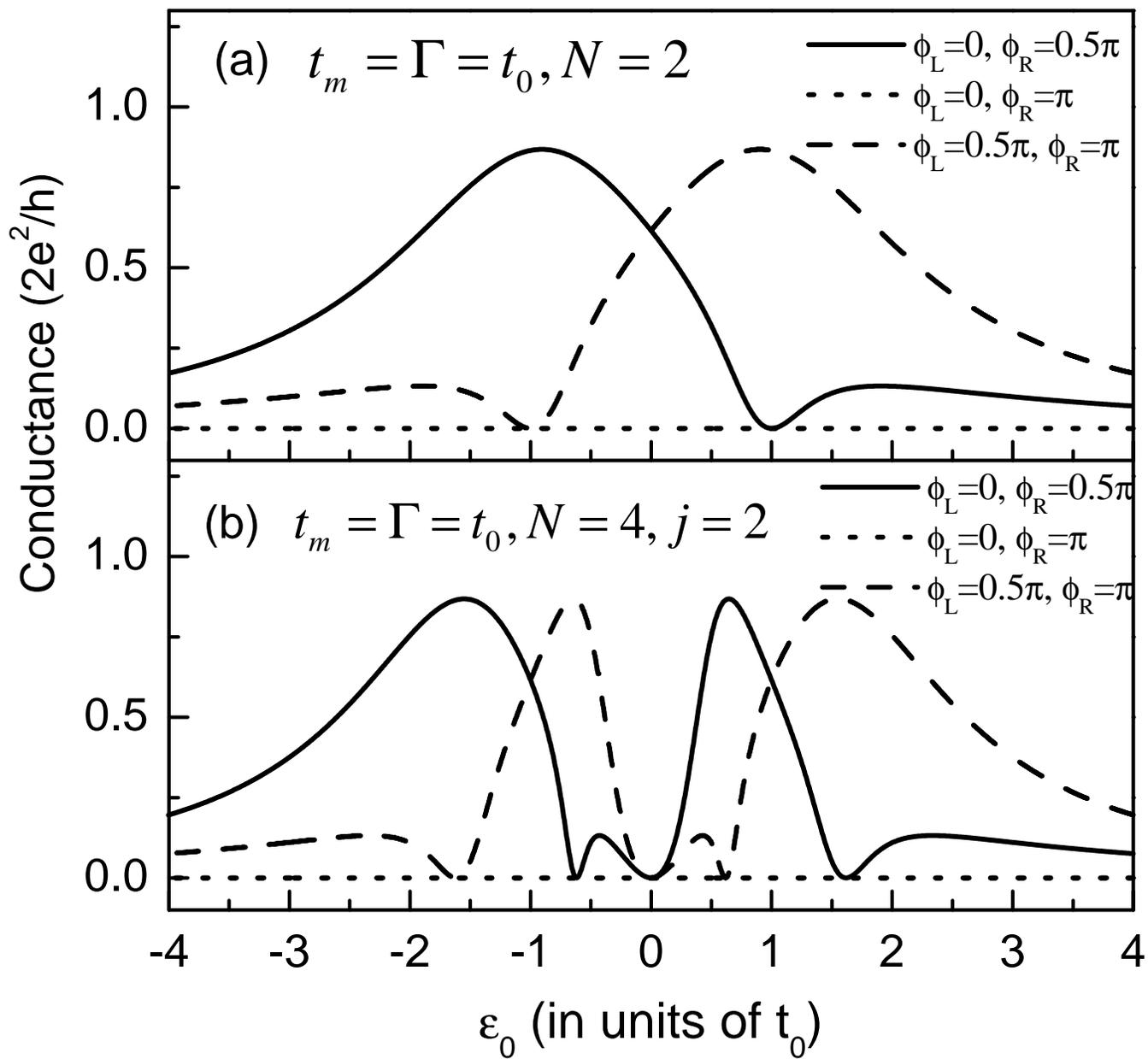
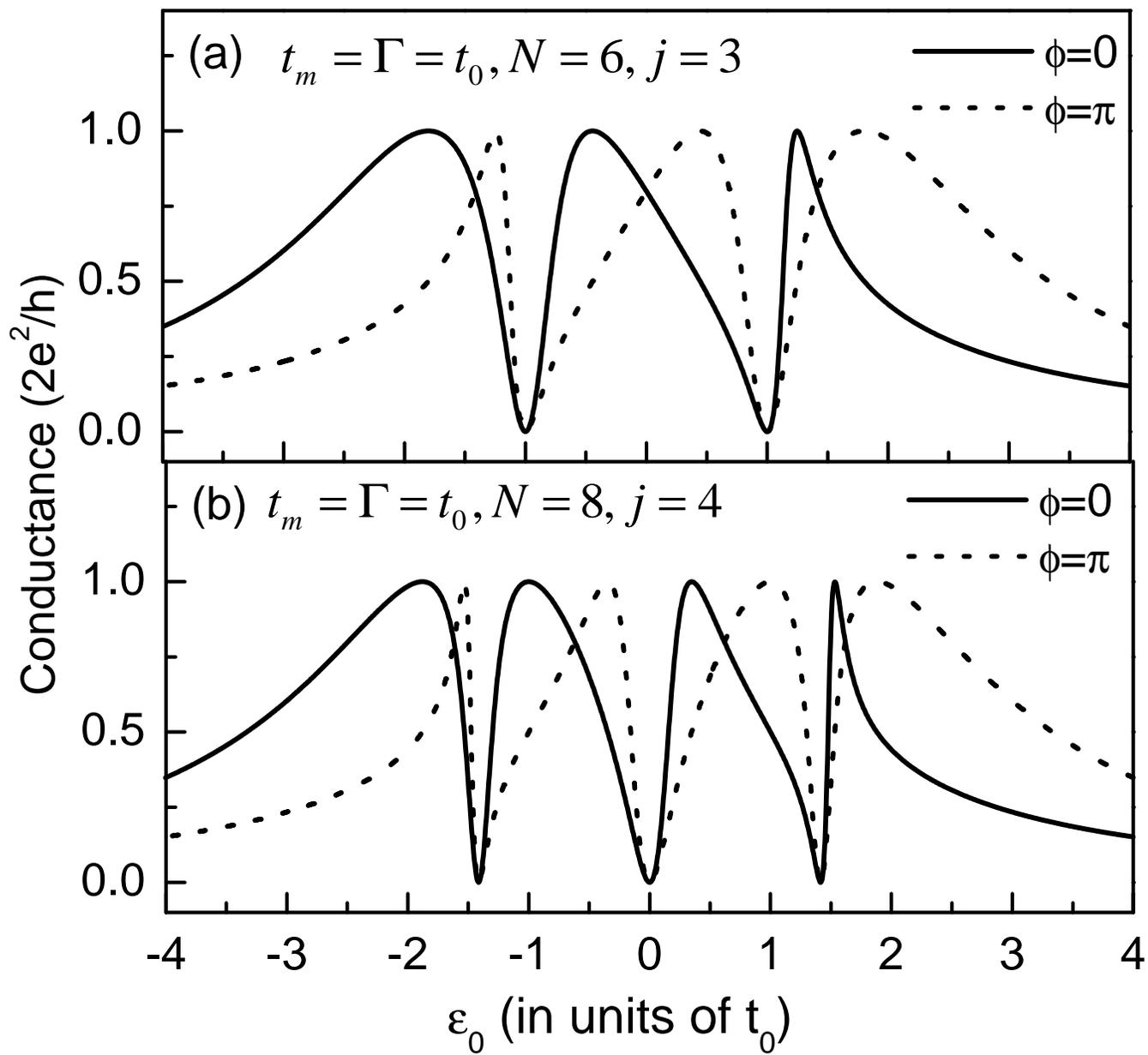
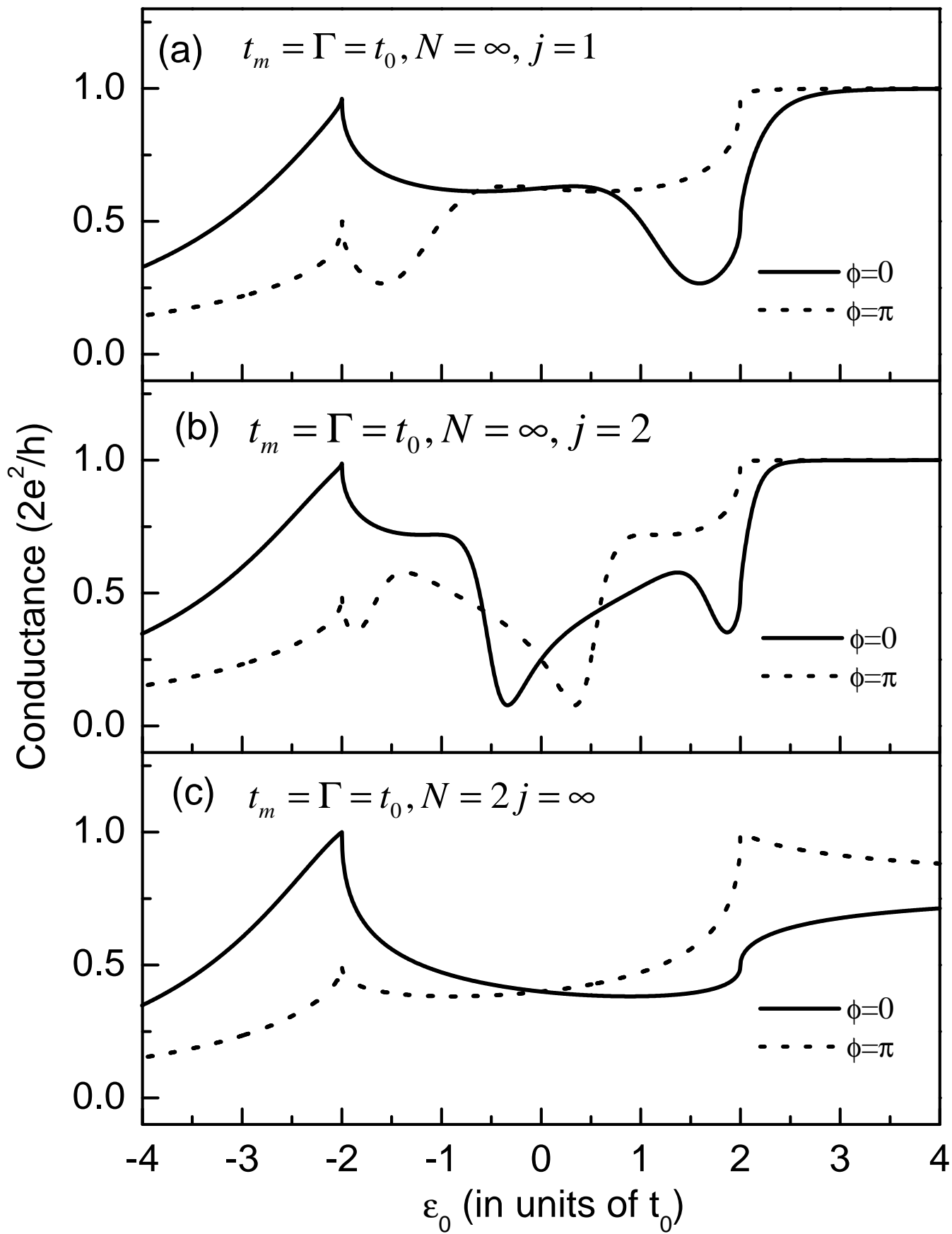
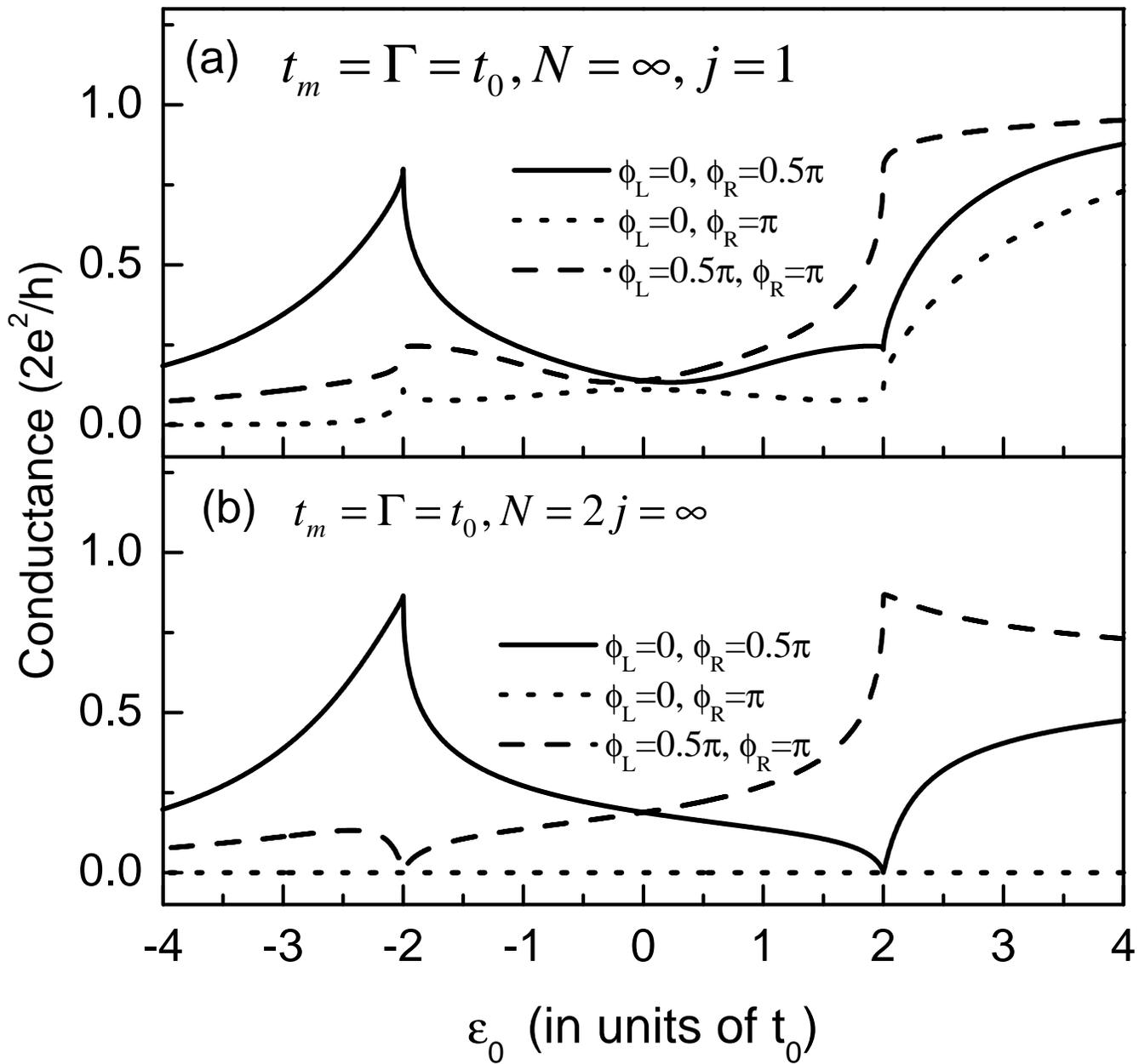
\end{document}